\documentclass[12pt]{iopart}

\usepackage{graphicx}% Include figure files
\usepackage{dcolumn}% Align table columns on decimal point
\usepackage{bm}% bold math
\usepackage{hyperref}% add hypertext capabilities

\usepackage{color}% Farbliche Markierungen von Textänderungen
\usepackage{ulem}% Zum Durchstreichen von alten Text

\usepackage{cite}

\begin{document}

%\title[Impact of magnetic frustration and site randomness on mK ADR]{Impact of magnetic frustration and site randomness on millikelvin adiabatic demagnetization refrigeration performance of rare-earth oxides}
\title[Utilizing frustration in Gd- and Yb-based oxides for mK ADR]{Utilizing frustration in Gd- and Yb-based oxides for milli-Kelvin adiabatic demagnetization refrigeration}
\author{Tim Treu, Marvin Klinger, Noah Oefele, Prachi Telang, Anton Jesche and Philipp Gegenwart}
\address{Experimental Physics VI, Center for Electronic Correlations and Magnetism, Institute of Physics, University of Augsburg, 86159 Augsburg, Germany}
\ead{philipp.gegenwart@physik.uni-augsburg.de}

\begin{abstract}
The manifold of energetically degenerate configurations arising from competing interactions in frustrated magnets gives rise to an enhanced entropy at lowest temperatures, which can be utilized for adiabatic demagnetization refrigeration (ADR). We review structural and magnetic properties of various Yb- and Gd-based oxides featuring frustration related to different triangular moment configurations and (in some cases) structural randomness. In comparison to paramagnetic hydrated salts, which have traditionally been employed for mK-ADR, these novel ADR materials enable cooling to temperatures several times lower than the magnetic interaction strength, significantly enhancing the entropy density and cooling power at a given target temperature. A further advantage is their chemical stability, allowing for a much simpler ADR pill design and ultra-high vacuum applications. For the temperature range between 0.03 and 2~K, a systematic comparison of the field-induced entropy density change is provided, that illustrates the advantages of frustrated magnets for low-temperature ADR.
%Quantum sciences and technologies typically require temperatures in the low Kelvin (K) or even milli-Kelvin (mK) range. Adiabatic demagnetization refrigeration (ADR) offers a sustainable and up-scalable alternative to dilution refrigeration with $^3$He/$^4$He mixtures. We review the suitability of recently investigated geometrically frustrated Yb- and Gd based borates, diphosphates, germanites and gallium garnets for mK-ADR. Compared to paramagnetic hydrated salts, traditionally used for mK-ADR, with crystal water separating the magnetic ions, magnetic frustration and randomness in these new ADR materials enables cooling to temperatures several times lower than the magnetic interaction strength, significantly enhancing the entropy density and cooling power. A further advantage is their chemical stability which enables a much simpler ADR pill design and also allows ultra-high-vacuum applications. ADR performance of various Yb- and Gd-based materials with different degree of geometrical frustration and randomness is analyzed for different temperature regimes.
\end{abstract}
%\keywords{adiabatic demagnetization refrigeration, sub-Kelvin temperature, magnetocaloric effect, magnetic order}
\submitto{\JPCM}
\maketitle
%\ioptwocol

\section{Introduction}\label{sec:introduction}

To study and apply the effects of quantum phenomena, low temperatures are needed to suppress thermal fluctuations and enable phase coherence. The investigation of matter close to absolute zero has led to important discoveries such as the quantum Hall effect, superfluidity and superconductivity. Ongoing research and recent developments in emerging fields such as quantum sensors and computers, and the desire to operate them on an industrial scale, have dramatically increased the demand for achieving stable thermal environments at very low temperatures. Adiabatic demagnetisation refrigeration (ADR) is suitable for achieving ultra-low temperatures.
In 1926 Debye and Giauque independently proposed that the magnetic disorder entropy of electronic magnetic moments in paramagnets could be used for cooling \cite{Debye1926}\cite{Giauque1927}. Seven years later, this proposal was realized when Giauque, together with MacDougall (0.52~K) and de~Haas, Wiersma, and Kramers (0.27~K) were the first to achieve temperatures significantly below 1 K by demagnetising paramagnetic salts \cite{Giauque1933, Haas1933}. Over time, this technique has been largely superseded by $^3$He-$^4$He dilution refrigeration, which has the significant advantage of being a continuous refrigeration method and providing higher cooling power \cite{Zu2022}. However, this method relies on the scarce and geopolitically problematic resource $^3$He, whose ever decreasing availability leads to price jumps and dependencies, making dilution refrigerators very costly to produce and operate \cite{Shea2010, Niechcial2020}. In addition, their design is very complex and large. Furthermore, they are not applicable in the emerging field of aerospace applications. A sustainable, helium-free alternative is currently being sought to meet the growing demand for low temperatures, leading to a revival of ADR with new theoretical concepts and new materials beyond the class of paramagnetic salts~\cite{Zhitomirsky2003,Hu2008,Evangelisti2011,Wolf2011,Wolf2014,Gruner2014,Jang2015,Tokiwa2016,
Wolf2016,Baniodeh2018,Gastaldo2019,Sereni2020,Shimura2022,Gruner2024}. In this paper, we discuss frustrated magnets as game changers for advanced mK refrigeration by ADR. 

\begin{figure}[t]
\centering
\includegraphics[width=0.4\textwidth]{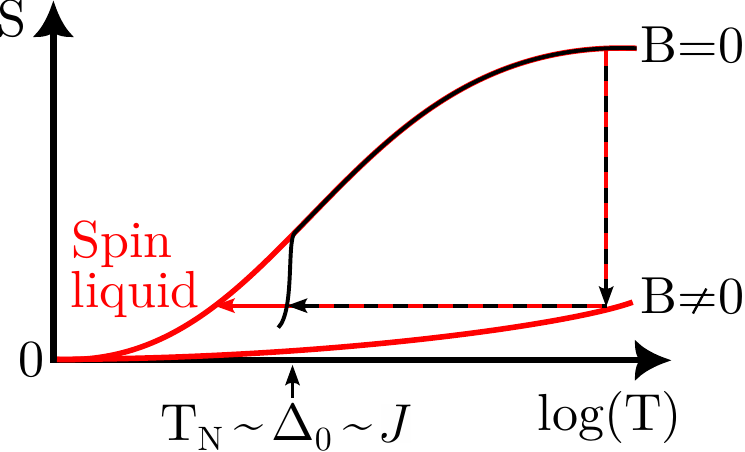}
\caption{Entropy curves at zero and a finite magnetic field as a function of temperature for conventional paramagnets (black) and a system with suppressed magnetic order (red)  \cite{Tokiwa2021}. $J$ and $\Delta_0$ denote the magnetic interaction and energy level splitting in zero field, respectively, which for a non-frustrated magnet determine the ordering temperature T$_\mathrm{N}$.}
\label{fig:MCE}
\end{figure}

In a typical ADR setup to attain sub-Kelvin temperatures, the material is precooled to a temperature in the range of 2 to 10 K, typically by a pumped $^4$He bath or a pulse tube cooler, while a magnetic field of a few T is applied to align the magnetic moments and thereby reduce the magnetic entropy to a minimum (vertical arrow in figure~\ref{fig:MCE}). Once thermal equilibrium is reached, the thermal contact to the heat bath is broken by opening a heat switch. Subsequently, the material is demagnetized under almost adiabatic conditions. This requires the temperature to be lowered when the field is driven back to zero (horizontal arrow in figure~\ref{fig:MCE}). Due to unavoidable external heat leakage the entropy eventually rises along the zero-field curve until its cooling power is exhausted~\cite{Pobell1996}. Unlike $^3$He-$^4$He dilution refrigeration, ADR is a one-shot cooling technique. Recent developments in multi-stage and continuous ADR setups may overcome this problem and potentially make ADR a dominant cooling technique in the mK range \cite{Shirron2004, Bartlett2010, Tuttle2017}.

Prerequisite for ADR are magnetic materials with a large magnetocaloric effect (MCE) which generate significant temperature changes when exposed to an external field. The ideal refrigerant would be a perfect paramagnet with negligible magnetic interactions, thus maintaining maximal zero-field entropy down to the lowest temperatures. However, in real materials the presence of finite interactions between the moments is inevitable, always driving the zero-field entropy towards zero. Therefore attempts at attaining the lowest possible ADR temperatures mainly followed the strategy to minimize the magnetic interaction by keeping the magnetic moments at very large distances~\cite{Wikus2014}.

%This leads to a tiny Zeeman splitting and magnetic order on the same energy scale ($\Delta_{\mathrm{0}} \sim$ J). Below the magnetic order temperature T$_{\mathrm{N}}$ the entropy abruptly drops to zero, limiting the final temperature T$_{\mathrm{N}} \sim \Delta_{\mathrm{0}} \sim$ J that can be reached by ADR with this material (see figure \ref{fig:MCE}). Consequently, the isothermal entropy change between H = 0 and H $\neq$ 0 is the key driving force for ADR. It is clear that ions with a large total angular momentum J will have a large entropy contribution, since the maximum magnetic entropy S$_{\mathrm{GS}}$ = Rln(2J+1) can be used for cooling. \\

Water-containing paramagnetic salts such as cerium magnesium nitrate (CMN) and ferric ammonium alum (FAA) were common early refrigerants and are still used in certain applications~\cite{Pobell1996, Fisher1973, Vilches1966}. These materials have very low ordering temperatures due to weak interactions caused by large separation between the magnetic ions, realized by incorporated water molecules. However, they suffer from several drawbacks as they can dehydrate and therefore decompose in vacuum or by already mild heating. ADR pills of these materials are difficult to prepare since the paramagnetic salts need to be grown in a dense mesh of metallic wires for thermalization and vacuum sealed to avoid degradation under evacuation~\cite{Bartlett2014}. Still such pills cannot be used for milli-Kelvin ultra-high vacuum (UHV) applications which require a heating of the ADR pill, as this leads to dehydration of the salts. For the oxide based magnets discussed in this review, mixing with fine silver powder and sintering provides simple and easy thermal contact~\cite{Tokiwa2021}. Due to the large interatomic distance of hydrated paramagnetic salts, their volumetric magnetic entropy density is rather low, giving rise to a reduced cooling power. This is particularly important where space is limited, e.g. in a cryostat or in aerospace applications. Generally magnetic materials with a higher density of magnetic ions exhibit stronger exchange interactions, resulting in higher ordering temperatures and thus preventing cooling to lower temperatures~\cite{Wikus2014}. Competing interactions, leading to frustration or even quantum criticality can reduce the ordering temperatures compared to ordinary magnets, providing a solution to this dilemma.

\begin{figure*}[t]
\centering
\includegraphics[width=0.99\textwidth]{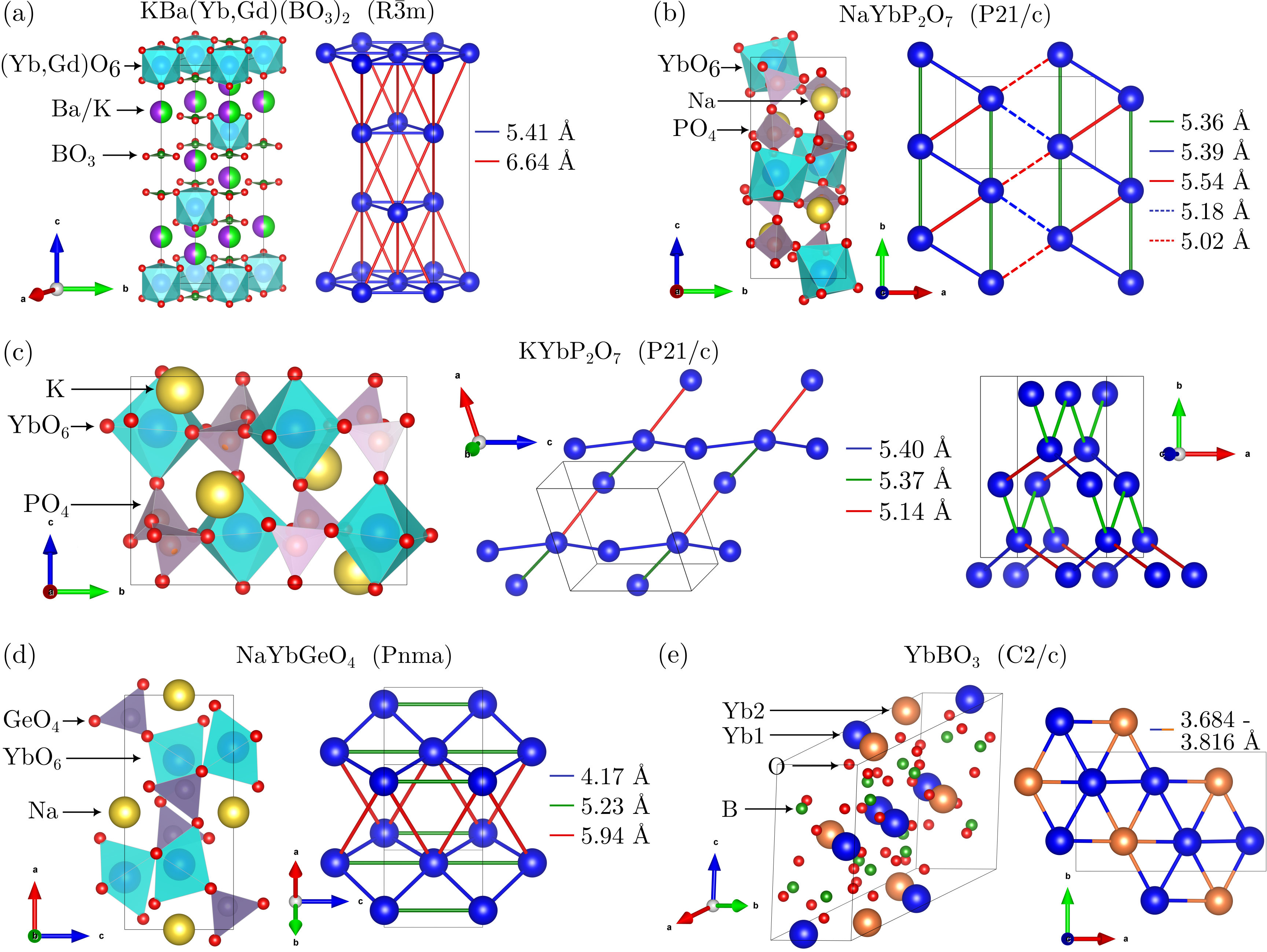}
\caption{Crystal structures and lattice geometries of the magnetic rare earth ions for various mK ADR materials. All structures have been drawn using Vesta software \cite{Vesta}.}
\label{fig:crystal_structures}
\end{figure*}

Figure~\ref{fig:MCE} compares schematically the temperature dependence of the entropy for the unfrustrated (black) and frustrated (red) cases. In the former, the ordering temperature is related to the magnetic interaction energy $J$ and the respective level splitting $\Delta_0$ in zero external field is due to the internal field related to magnetic ordering. This energy limits the minimal achievable temperature for ADR. On the other hand, a frustrated magnet remains disordered below the temperature of the dominating magnetic interaction $J$, entering a spin liquid state upon cooling \cite{Li2020}. Compared to the ordered state, the manifold of soft modes resulting from the correlations in the spin liquid state lead to an enhanced zero-field entropy \cite{Zhitomirsky2003}. This allows lower ADR final temperatures compared to an unfrustrated system with similar magnetic density. Respectively an unfrustrated magnet reaching a similar ADR final temperature than the frustrated magnet would have to have a lower magnetic coupling strength and thus a lower magnetic ion density. We note that structural disorder can smear out magnetic phase transitions and suppress long range order, leading to spin glassy behavior. In this case, the sharp drop in the black curve is broadened and cooling to lower temperatures is possible, though the larger the size of slowly fluctuating clusters, the smaller their contribution to the total magnetic entropy.
%
%Furthermore the size $S$ of the spins, which determines their molar entropy $R\log(2S+1)$ should be as high as possible. Additionally, earlier theoretical work has shown that in frustrated magnets the magnetocaloric effect below the saturation field is enhanced by the existence of a macroscopic number of soft mode excitations arising from classical ground state degeneracy \cite{Zhitomirsky2003}. However, it has been experimentally observed that the contribution to the magnetocaloric entropy change is more dominant for the nearest neighbour coupling than for the number of soft mode excitations in the liquid He regime \cite{Koskelo2023}.

The probably best studied frustrated magnet for low-T ADR applications is the garnet Gd$_3$Ga$_5$O$_{12}$ (GGG), in which Gd$^{3+}$ ions form two interpenetrating rings of corner shared triangles \cite{Koskelo2023}. In this material, the frustration is sufficiently strong to prevent long-range magnetic order down to 27~mK~\cite{Bonville2004}. Zero-field specific heat shows only a broad hump around 0.8 K~\cite{Onn1967}, while a field-induced antiferromagnetic transition was observed near 0.35~K at 1~T~\cite{Schiffer1994}. GGG has a very high volumetric entropy density and is a suitable ADR material at temperatures of order 1~K~\cite{Kleinhans2023}. Compared to GGG with $S=7/2$, isostructural Yb$_3$Ga$_5$O$_{12}$ (YbGG) with effective spin 1/2 moments arising from the lowest Kramers doublet features a lower magnetic coupling, resulting in a broad specific heat hump near 0.2 K and can be utilized for ADR down to 0.1 K~\cite{PaixaoBrasiliano2020}.
More recent promising ADR candidates are intermetallic systems like YbCu$_4$Ni \cite{Shimura2022, Sereni2018} and YbPt$_{\mathrm{2}}$(Sn,In) \cite{Jang2015,Gastaldo2019}, which benefit from the enhanced thermal conductivity facilitated by conduction electrons, while maintaining a relatively high volumetric entropy density. The system YbNi$_{\mathrm{1.6}}$Sn \cite{Gruner2024} is the most promising, exhibiting minimal Kondo and Ruderman-Kittel-Kasuya-Yosida (RKKY) interactions, enabling refrigeration down to about 120~mK.

In this review, we focus our attention on a couple of most recently investigated frustrated Ytterbium and Gadolinium oxides for the milli-Kelvin ADR application, whose crystal structures are displayed in figure~\ref{fig:crystal_structures}.
%In the following chapters, the different crystal structures, magnetic and thermodynamic properties are discussed, before the ADR performance of these new materials is compared with a broader range of other ADR materials. 
Here we only provide a first overview, while details will be discussed afterwards in separate chapters. In KBaYb(BO$_{\mathrm{3}}$)$_{\mathrm{2}}$ magnetic order is suppressed by magnetic frustration and site randomness, resulting in a high entropy density and low transition temperature ($<$ 10 mK)~\cite{Tokiwa2021}. The isostructural sister compound KBaGd(BO$_{\mathrm{3}}$)$_{\mathrm{2}}$ has a three times higher magnetic entropy density due the Gd $S=7/2$ spins~\cite{Jesche2023}. The better refrigeration performance is limited by a higher ordering temperature of 263 mK, making it suitable as excellent pre-cooling refrigerant in a multi-stage ADR setup~\cite{Shirron2014}.
%At the pre-cooling stage the electrical wires to the main cooling stage, the radiation shield and other parts of the setup can be thermally anchored, thereby reducing the heat load on the main stage. This can significantly increase the operating temperature range, efficiency and cooling capacity \cite{Shirron2014}.
Other new efficient frustrated mK-ADR materials are the diphosphates KYbP$_{\mathrm{2}}$O$_{\mathrm{7}}$ and NaYbP$_{\mathrm{2}}$O$_{\mathrm{7}}$. While the former reaches even lower temperatures than KBaYb(BO$_{\mathrm{3}}$)$_{\mathrm{2}}$ in an ADR experiment under comparable conditions, the latter provides a longer hold time below 2 K \cite{Arjun2023Phosphate}. These compounds may be preferable for specific temperature profiles. The importance of geometrical frustration for ADR is exemplified by the comparison of ADR properties of a respective non-frustrated system NaYbGeO$_{\mathrm{4}}$~\cite{Arjun2023} in section 4. After discussion of another related triangular quantum magnet YbBO$_{\mathrm{3}}$~\cite{Sala2023}, we end the paper with  a systematic comparison of the ADR performance of these systems, with respect to other known sub-Kelvin ADR materials in section 6 before ending with summary and conclusion.

\section{KBa(Yb,Gd)(BO$_{\mathrm{3}}$)$_{\mathrm{2}}$}

\begin{figure}[t]
\centering
\includegraphics[width=0.8\textwidth]{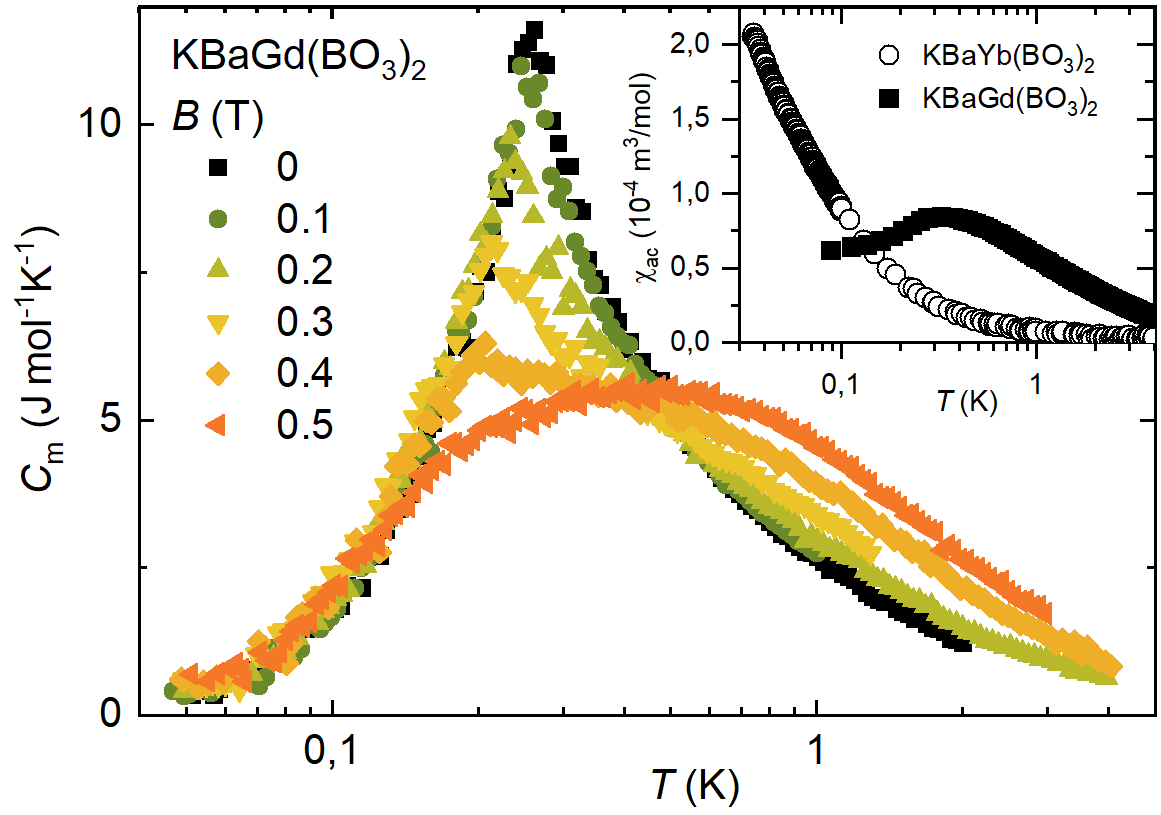}
\caption{Magnetic contribution to the specific heat of KBaGd(BO$_{\mathrm{3}}$)$_{\mathrm{2}}$ in zero and small magnetic field~\cite{Jesche2023}. The inset displays the magnetic ac susceptibility ~\cite{Xac} of KBaGd(BO$_{\mathrm{3}}$)$_{\mathrm{2}}$ (closed squares) in comparison with KBaYb(BO$_{\mathrm{3}}$)$_{\mathrm{2}}$ (open circles).}
\label{fig:HC_comparison_borates}
\end{figure}

\begin{figure}[t]
\centering
\includegraphics[width=0.7\textwidth]{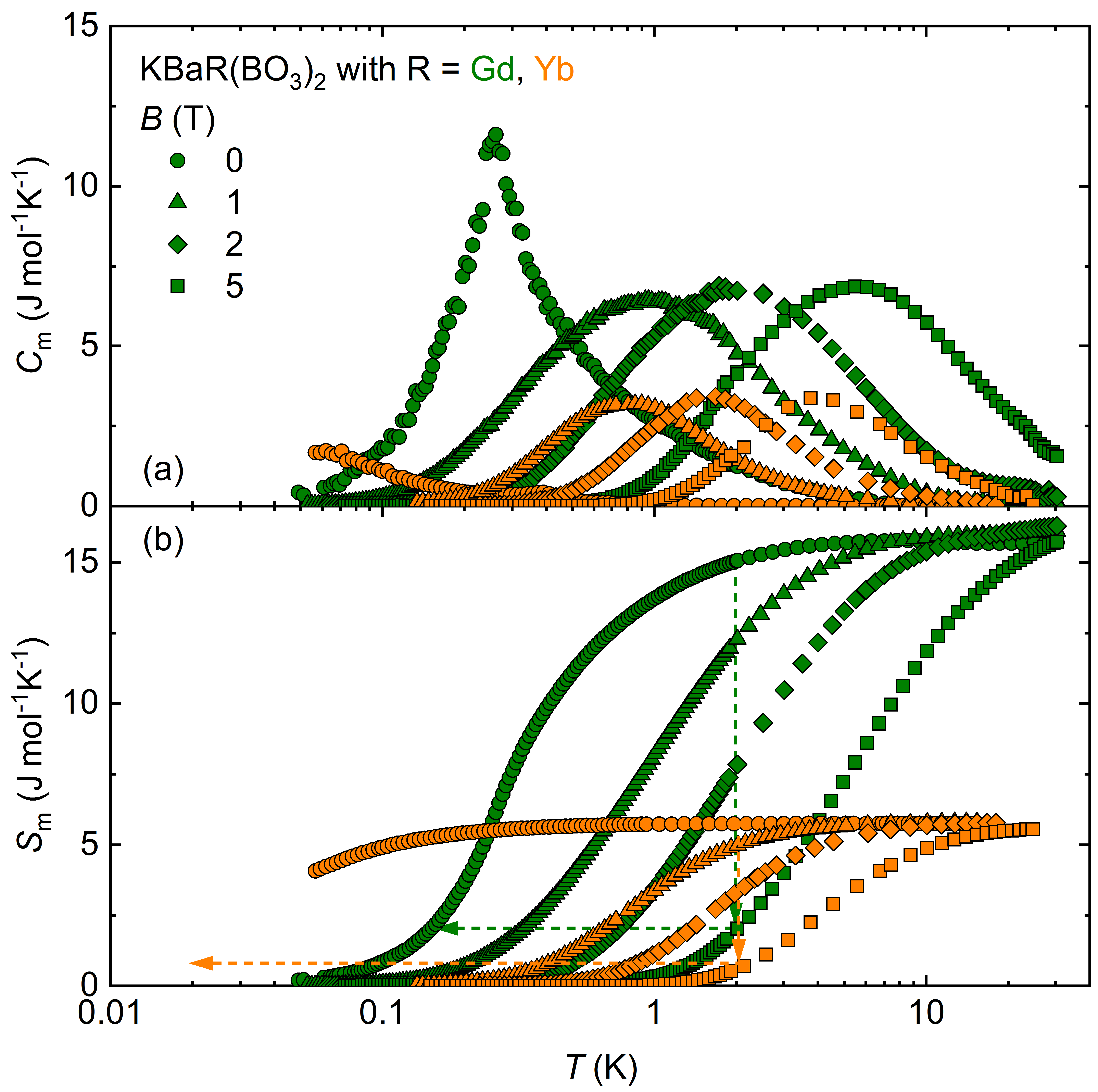}
\caption{Magnetic contribution to the specific heat (a) and entropy (b) of KBaYb(BO$_3$)$_2$ (orange) \cite{Tokiwa2021} and KBaGd(BO$_3$)$_2$ (olive) \cite{Jesche2023} on a logarithmic temperature scale for various applied magnetic fields. The arrows in (b) indicate the entropy reduction under isothermal magnetization to 5~T and subsequent temperature reduction under adiabatic demagnetization.}
\label{Figure4}
\end{figure}

KBaYb(BO$_{\mathrm{3}}$)$_{\mathrm{2}}$ and KBaGd(BO$_{\mathrm{3}}$)$_{\mathrm{2}}$ both crystallize in a rhombohedral lattice with the R$\overline{\mathrm 3}$m space group (see figure~\ref{fig:crystal_structures}(a)) \cite{Tokiwa2021, Jesche2023, Guo2019}. Here (Yb/Gd)O$_{\mathrm{6}}$-octahedra form a triangular lattice in the ab-plane, separated by BO$_{\mathrm{3}}$ triangles. Adjacent triangular layers along the c-axis are separated by positively charged K$^+$ and Ba$^{2+}$ ions and are shifted laterally with respect to each other. This shift causes additional frustration. The site randomness of the K$^+$ and Ba$^{2+}$ ions leads to non-uniform charge distribution and thus to uneven electric fields acting on the magnetic rare earth ions. Triangular lattice YbMgGaO$_4$ with statistical site randomness of Mg$^{2+}$ and Ga$^{3+}$ adjacent to Yb$^{3+}$ shows broadened crystal electric field excitations and a randomized magnetic exchange~\cite{Li2020}. Therefore the site randomness in the borates may also influence their magnetic properties. Indeed, the AFM order found at 0.26~K in KBaGd(BO$_{\mathrm{3}}$)$_{\mathrm{2}}$~\cite{Jesche2023} manifests in broadened peaks of the zero-field specific heat and the magnetic ac susceptibility~\cite{Xac}, cf. figure~\ref{fig:HC_comparison_borates}, indicative of significant rounding of the phase transition. Furthermore, only about 1/3 of the total entropy is recovered at this temperature. Application of fields below 0.5~T shifts the peak towards lower temperatures, consistent with AFM order, while in addition a shoulder emerges. Beyond 0.4~T the latter develops into the broadened Schottky-type hump, characteristic for the Zeeman splitting, that shifts towards higher temperatures with increasing field (see figure~\ref{Figure4}). Comparison of $C_m(T)$ at fields between 1 and 5 T with the expectation from the temperature derivative of 
$E_{\rm mag}=-\mu_{\rm sat} B_J B$ ($B_J$: Brillouin function) yields further evidence for AFM interactions between the Gd$^{3+}$ spins in KBaGd(BO$_{\mathrm{3}}$)$_{\mathrm{2}}$~\cite{Jesche2023}.

As shown in figure~\ref{Figure4} and the inset of figure~\ref{fig:HC_comparison_borates}, the zero-field specific heat and the magnetic ac susceptibility of isostructural KBaYb(BO$_{\mathrm{3}}$)$_{\mathrm{2}}$ show no sign of a magnetic transition down to 0.03~K. In fact a very low AFM ordering temperature of 9 mK can be deduced by assuming that the N\'{e}el temperature scales with the square of the moment size~\cite{Jesche2023}.

Next, we compare the magnetic specific heat and entropy of the Gd- and Yb-borate materials in differing fields up to 5~T, displayed in figure~\ref{Figure4}. Note, that the (field-independent) lattice contribution, which is negligibly small below 2~K, has been subtracted, yielding a high-temperature saturation of the entropy at $R\ln(2S+1)$. Compared to the Yb-system, the $S=7/2$ state in the Gd-system has a three times larger magnetic entropy and undergoes a stronger Zeeman-splitting, indicated by the stronger shift of the Schottky anomaly in the magnetic specific heat (cf. green versus orange curves). The vertical arrows in panel (b) indicate the entropy reduction to about $10\%$ of its value in zero field upon isothermal magnetization from 0 to 5~T field at a temperature of 2~K for the two materials. The subsequent demagnetization from the initial field $B_{\rm i}=5$~T under perfect adiabatic conditions would reduce the temperature to $\sim 150$~mK and to below 30~mK for the Gd- and Yb-materials, respectively.

\begin{figure}[t]
\centering
\includegraphics[width=0.99\textwidth]{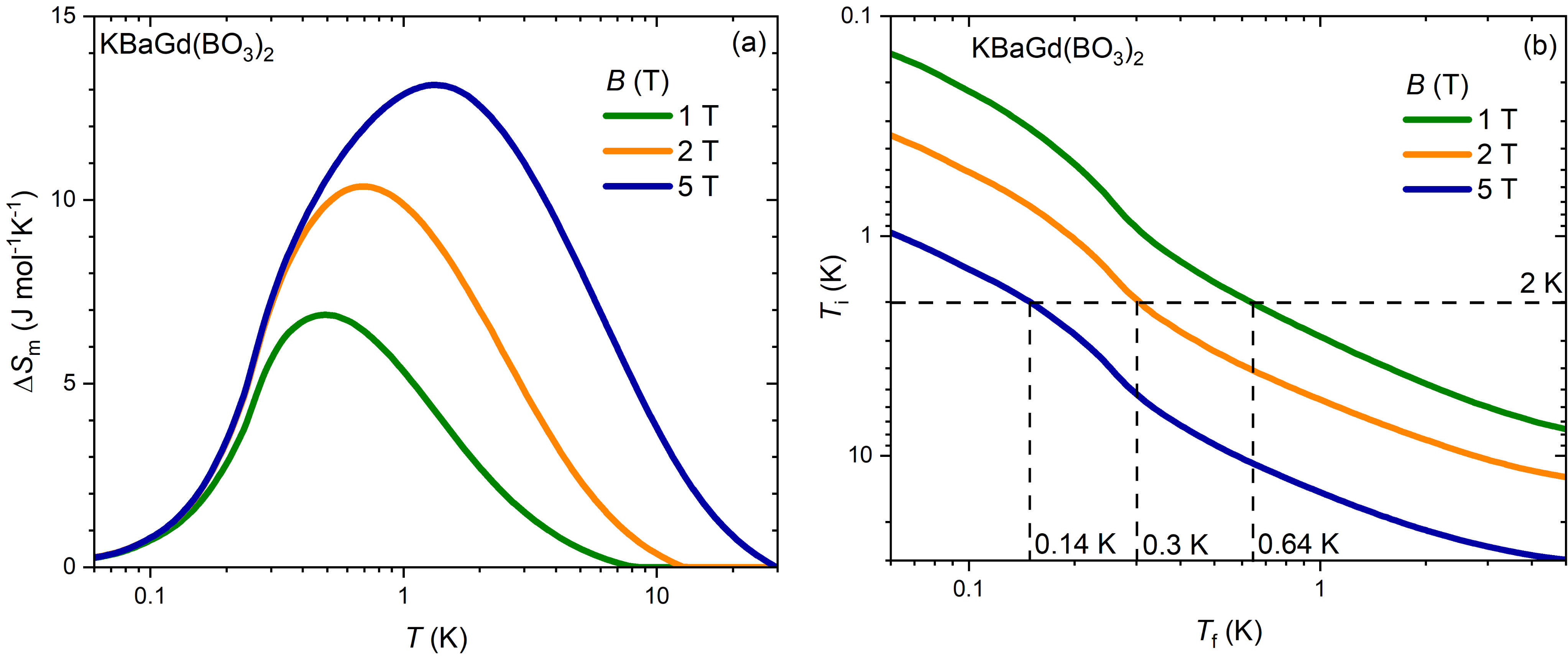}
\caption{Magnetic molar entropy increment $\Delta S_{\rm m}(T)=S_{\rm m}(T,0)-S_{\rm m}(T,B)$ at $B=1$, $2$ and $5$~T for KBaGd(BO$_{\mathrm{3}}$)$_{\mathrm{2}}$ (a). Respective ADR initial and final temperatures as $T_i$ vs $T_f$, calculated from the magnetic entropy (b). Note the reversed scale for $T_i$.}
\label{Figure5}
\end{figure}

% Fitting gives a Curie-Weiss temperature of -38 mK, which is significantly lower than the value of -60 mK achieved in measurements done with a $^3$He option. The magnetic interaction is therefore even weaker than previously assumed. The large cooling capabilities arise from the fact that this weak interaction is of the same order of magnitude as the dipolar coupling. Not only is the entropy density higher than that of the paramagnetic salts (see section \ref{sec:comparison}), but the absence of crystalline water makes the compound chemically stable in vacuum and at high temperatures, making it suitable for use in UHV applications as an excellent refrigerant. Good thermal contact can be achieved by mixing the sample powder with fine silver powder and pressing it into pellets, further simplifying the application. The low cost and ease of production underlines its high potential in practical implementation.
 
To illustrate the ADR suitable temperature range of a certain material it is appropriate to consider the change of molar magnetic entropy between zero and finite magnetic field. Figure \ref{Figure5}(a) displays $\Delta S_{\rm m}(T)=S_{\rm m}(T,0)-S_{\rm m}(T,B)$ for KBaGd(BO$_{\mathrm{3}}$)$_{\mathrm{2}}$ for $B=1$, 2 and 5~T. It is obvious that increasing the ADR initial field $B$ enhances the maximal entropy increment and shifts maxima temperatures in $\Delta S_{\rm m}(T)$ towards higher values. At the largest considered initial field of 5~T the maximal entropy increment of  76\% of $R\ln 8$ can utilized when working at an initial ADR temperature of
1.5~K. Larger fields would be required to shift the maximum above 2~K. Estimation of isothermal entropy change is relevant for the determination of the optimal operating temperature for the ADR material in a multi-stage continuous ADR cooler that is kept at a constant temperature. If the entropy difference is not normalized per mol but scaled as volumetric density, such representation allows to compare the ADR suitability of different materials over wide temperature ranges. Respective comparative plots follow in the last section of this review. On the other hand, in a single stage one-shot process, the relation between the initial and final temperature when demagnetizing from $B_{\rm i}$ to zero field is of further relevance. For KBaGd(BO$_{\mathrm{3}}$)$_{\mathrm{2}}$ this relation is displayed in panel (b) of figure \ref{Figure5} for three different initial fields 1, 2 and 5~T. The dashed lines indicate the obtained end temperatures for differing starting fields at an initial temperature of 2~K. This plot thus allows to directly determine the achievable final temperature $T_{\rm f}$ for a given initial temperature and field.

The suitability of KBaYb(BO$_{\mathrm{3}}$)$_{\mathrm{2}}$ and KBaGd(BO$_{\mathrm{3}}$)$_{\mathrm{2}}$ for ADR has been directly tested in a home-made setup mounted on the Quantum Design physical property measurement system (PPMS) puck~\cite{Tokiwa2021,Jesche2023}. Pressed pellets with 15~mm diameter of the borates with admixture of fine silver powder for thermalization were mounted on a thermally insulating holder, shielded from radiation by a brass cap. Adiabaticity was reached by attaining high vacuum inside the sample chamber. Starting at 2~K in 5~T, final temperatures $T_{\rm f}$ of 40~mK~\cite{Tokiwa2021} and $\sim 0.14$~K~\cite{Jesche2023} were obtained for the Yb- and Gd-system, respectively, with an order of magnitude slower warmup back to 2~K in the latter, due to its larger magnetic specific heat. Of course such a simple setup is far from adiabatic. As discussed in more detail below, the warmup curve can be used to estimate the specific heat of the ADR material and the heat load of the ADR setup, which has been below $1~\mu$W in these experiments~\cite{Jesche2023}. Further ADR testing of KBaYb(BO$_{\mathrm{3}}$)$_{\mathrm{2}}$ under improved adiabatic conditions in a dilution refrigerator revealed even a minimal temperature below 20~mK for the same initial parameters~\cite{Tokiwa2021}. Note that the admixture of fine Ag powder to the pressed ADR pellets ensures excellent thermal coupling even at these extremely low temperatures. Such pellets are mechanically stable and, in contrast to pellets with hydrated paramagnetic salts, can be baked out for applications in ultra-high-vacuum such as the recently constructed mK-scanning probe microscope~\cite{Esat2021}.

%the adiabatic temperature changes $\Delta$T$_{\mathrm{ad}}$ for different fields. For reference, the expected final temperatures T$_{\mathrm{f}}$ are given for an initial temperature T$_{\mathrm{i}}$ = 2 K. Values for T$_{\mathrm{f}}$ = 0.14, 0.3 and 0.64 respectively are obtained for demagnetisation from 5, 2 and 1 T to 0 T. This is in agreement with the lowest temperature measured of 122 mK in a commercial PPMS by starting demagnetisation at 2 K with 5 T \cite{Jesche2023}. It should be noted that the difficulty of achieving complete thermal insulation in a practical setup complicates the interpretation of the adiabatic temperature change. Overall, the high entropy density with limited end temperatures makes this compound excellent for the first cooling stage in a multi-stage setup. To achieve ultra-low temperatures other compounds like KBaYb(BO$_{\mathrm{3}}$)$_{\mathrm{2}}$ can be used.
%
% In this case, only 76\% of the magnetic entropy can be used for cooling, indicating that higher fields are required to exploit the full potential. 

\section{(Na,K)YbP$_{\mathrm{2}}$O$_{\mathrm{7}}$}

Next, we briefly discuss the two diphosphates NaYbP$_{\mathrm{2}}$O$_{\mathrm{7}}$ and KYbP$_{\mathrm{2}}$O$_{\mathrm{7}}$, whose crystal structures are also displayed in figure~\ref{fig:crystal_structures}. Both crystallize in a monoclinic lattice with space group P21/c, but show different structure types \cite{Arjun2023Phosphate, Ferid2004, HorchaniNaifer2007}. In NaYbP$_{\mathrm{2}}$O$_{\mathrm{7}}$ the YbO$_{\mathrm{6}}$ octahedra form a triangular layer in the ab plane (see figure~\ref{fig:crystal_structures}(b)). However, they do not form an ideal equilateral triangular lattice, rather there are five distinct lengths between 5.02 and 5.36~$\mathring {\mathrm A}$. Along the c-axis these layers are separated by PO$_{\mathrm{4}}$ tetrahedra and Na ions. In contrast, in KYbP$_{\mathrm{2}}$O$_{\mathrm{7}}$, YbO$_{\mathrm{6}}$ octahedra are connected by PO$_{\mathrm{4}}$ tetrahedra and form jagged chains along the c-axis (see figure~\ref{fig:crystal_structures}(c)). These chains are connected by alternating chains in the ac plane, forming a three-dimensional distorted honeycomb-like structure. Altogether these two materials also feature geometrical frustration, though in contrast to the borates they feature structural distortions but no site randomness. Similar to KBaYb(BO$_{\mathrm{3}}$)$_{\mathrm{2}}$, paramagnetic behavior is found in the magnetic susceptibility down to 0.4~K with very low Curie-Weiss constants for both materials, characteristic of a very weak magnetic interaction~\cite{Arjun2023Phosphate}. Indeed the magnetic ion density is very similar to that of KBaYb(BO$_{\mathrm{3}}$)$_{\mathrm{2}}$. Relatedly, also the specific heat of both systems at 0.4 - 20~K in fields up to 5~T resembles very much the data shown in figure~\ref{Figure4}(a) for the Yb-borate. There are no investigations to below 40 mK yet, which could yield the magnetic ordering temperatures of these two materials. From the ADR performance tests in the PPMS any ordering above 40 mK can be excluded. Both materials allow the cooling to rather  similar final temperatures, but with fine differences. Compared to  KBaYb(BO$_{\mathrm{3}}$)$_{\mathrm{2}}$, KYbP$_{\mathrm{2}}$O$_{\mathrm{7}}$ reaches a 20\% lower end temperature (but 12.5\% shorter warmup time), while for NaYbP$_{\mathrm{2}}$O$_{\mathrm{7}}$ a similar end temperature is combined with 30\% longer hold time~\cite{Arjun2023Phosphate}. This may also be in line with the approximately 10\% lower magnetic entropy density of KYbP$_{\mathrm{2}}$O$_{\mathrm{7}}$ compared to the Na system and the borate (which have rather similar density), leading to even lower magnetic exchange and ordering temperatures. However, very low-temperature thermodynamic experiments are required to conclude about the ordering and amount of frustration in these materials.

\section{NaYbGeO$_{\mathrm{4}}$}

NaYbGeO$_{\mathrm{4}}$ crystallizes in an orthorhombic lattice with space group Pnma \cite{Arjun2023, EmirdagEanes2001}. As can be seen in figure~\ref{fig:crystal_structures}(d), the YbO$_{\mathrm{6}}$ octahedra are cornersharing, lying on the corners of a distorted square lattice in the bc plane. Along the a-axis they are separated by GeO$_{\mathrm{4}}$ tetrahedra and Na ions. Again, magnetization and specific heat data indicate weakly interacting effective spin 1/2 moments below 10 K, compatible with a crystal electric field Kramers ground state doublet. In the spin 1/2 two-dimensional frustrated square lattice or J$_{\mathrm{1}}$-J$_{\mathrm{2}}$ model, frustration can arise due to competing interactions such as $J_1$ between nearest neighbors along the edges and $J_2$ between next-nearest neighbors along the diagonal direction. If both $J_1$ and $J_2$ are AFM, frustration occurs, leading to a rich theoretical phase diagram with various ground states depending on $J_1/J_2$~\cite{Shannon2004}.

Low temperature fits of the inverse susceptibility yield a CW temperature of 15~mK, suggesting a very low ordering temperature. In fact, however an AFM phase transition is observed at 0.21~K in the specific heat calculated from the ADR warmup curve in the PPMS~\cite{Arjun2023}. Since the value of $\theta_{\mathrm{CW}}$ depends on the sum of the FM and AFM contributions, the fact that $\theta_{\mathrm{CW}}$ is much smaller than the ordering temperature can be explained by a superposition of AFM nearest neighbor and FM next-nearest neighbor exchange couplings on the square lattice. As a result, there appears no frustration as both first and second neighbor interactions are satisfied simultaneously in the AFM ground state. Due to the absence of frustration and the onset of magnetic order at 0.21~K, the lowest temperatures achievable with ADR are limited. Cooling below 150 mK is impossible, different to KBaYb(BO$_{\mathrm{3}}$)$_{\mathrm{2}}$ and (Na,K)YbP$_{\mathrm{2}}$O$_{\mathrm{7}}$, for which final temperatures below 50 mK were obtained (see section \ref{sec:comparison}). This demonstrates not only the importance but also the necessity of magnetic frustration in achieving low temperatures while maintaining a high volumetric entropy density.

\section{YbBO$_{\mathrm{3}}$}\label{sec:YbBO3}

Triangular YbBO$_{\mathrm{3}}$ has been the subject of several studies in recent years with conflicting results. It has a monoclinic structure with space group C2/c~\cite{Mukherjee2018,Sala2023}. Two distinct Yb sites are present, forming a triangular lattice in the ab plane with small deviations from the ideal equilateral shape (see figure~\ref{fig:crystal_structures}(e)). These deviations potentially alleviate the frustration expected for a canonical triangular lattice. The much smaller distance between the magnetic ions increases the interaction compared to KBaYb(BO$_{\mathrm{3}}$)$_{\mathrm{2}}$ and both (Na,K)YbP$_{\mathrm{2}}$O$_{\mathrm{7}}$. The Yb layers are separated by B$_{\mathrm{3}}$O$_{\mathrm{9}}$ units.

\begin{figure}[t]
\centering
\includegraphics[width=0.6\textwidth]{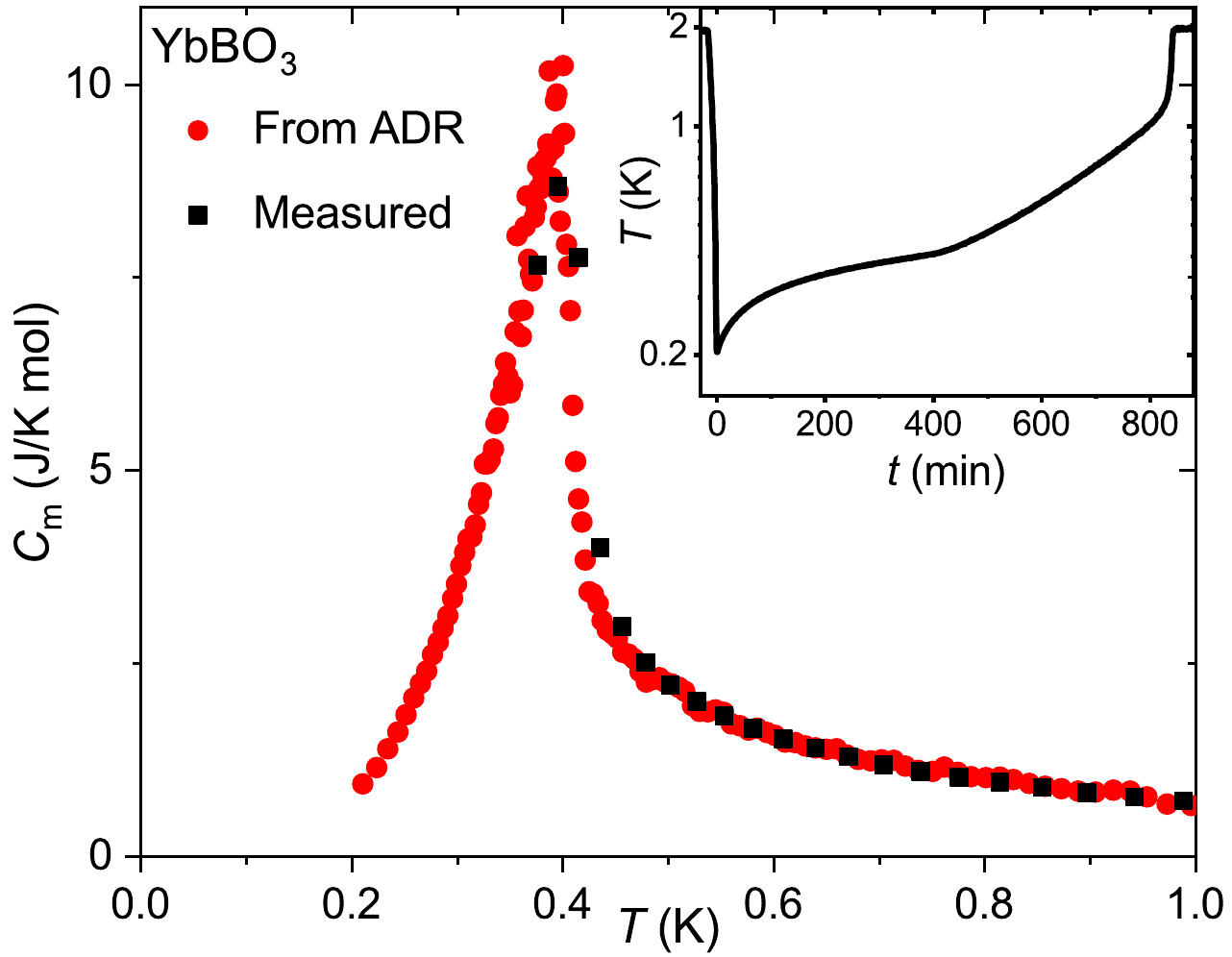}
\caption{Comparison of the magnetic heat capacity of the YbBO$_{\mathrm{3}}$ ADR pellet C$_{\mathrm{ADR}}$~=~$\dot{Q}$/$\dot{T}$ derived from the ADR experiment warming curve (inset) and the measured heat capacity. Good agreement was obtained for $\dot{Q}$ = 0.26 $\mu$W.}
\label{fig:YbBO3_HC}
\end{figure}

Low temperature measurements were first reported by Somesh {\it et al.} in early 2023~\cite{Somesh2023}. No phase transition or spin freezing was reported down to 20 mK using $\mu$-SR measurements on polycrystalline powder samples. This leads the authors to conclude a highly frustrated magnetic state in YbBO$_3$. On the other hand, Sala {\it et al.} later found a clear indication of AFM order at 0.4~K in specific heat measurements~\cite{Sala2023}. They have studied a pressed pellet with an equal mass of high purity Ag powder admixture. To investigate the ADR performance of YbBO$_3$, we investigated the ADR performance of pressed pellets from YbBO$_3$ powder from~\cite{Somesh2023} and our own synthesis, both mixed with fine Ag powder to improve thermal contact; both samples do show similar behavior. Again an initial temperature of 2~K and initial field of 5~T were chosen. A final temperature of 0.2~K has been achieved. The warm-up curve, collected over 14 hours allowed to estimate the heat capacity using the equation
%
%For the powder from this report, as well as our own synthesis, the ADR performance was measured on a pressed pellet with Ag powder admixture in the PPMS, cf. the inset of Fig.~\ref{fig:YbBO3_HC}. A final temperature of 0.2~K and a warm-up time over 14 hours was achieved. From this warming curve the heat capacity curve shown by the red dots was calculated using

\begin{equation}
\dot{Q} = C_{\mathrm{ADR}} \cdot \dot{T},
\label{eq:ADR}
\end{equation}

under the assumption of a temperature independent constant heat input. Comparison with directly measured specific heat data down to 0.4~K (in the PPMS) yields $\dot{Q}$ = 0.26 $\mu$W. Figure~\ref{fig:YbBO3_HC} displays the sharp AFM phase transition, calculated from the warmup curve shown in the inset, which is also in good agreement with the data from~\cite{Sala2023}. Note that the heat input in this experiment has been lower compared to the estimated values for KBaGd(BO$_{\mathrm{3}}$)$_{\mathrm{2}}$ (0.71 $\mu$W) \cite{Jesche2023} and NaYbGeO$_{\mathrm{4}}$ (0.58 $\mu$W) \cite{Arjun2023}, where the measurements were performed with the same setup in the same PPMS. This is due to the regeneration of the cryogenic pump, which creates a better vacuum and thus reduces the heat transfer by convection through residual He gas. This comparatively low heat leakage also explains the long warm-up time of YbBO$_{\mathrm{3}}$, since it scales with the heat transfer rate.

\section{Comparison}\label{sec:comparison}

\begin{table}
\caption{Comparison of key parameters of different mK ADR materials sorted by volumetric entropy density: T$_\mathrm{m}$ is the magnetic ordering temperature (in some cases only short-range), T$_\mathrm{min}$ is the minimum temperature attained, S$_\mathrm{GS}$ is the ground state multiplet entropy and R is the universal gas constant. The abbreviations stand for: CMN~=~Mg$_\mathrm{3}$Ce$_\mathrm{2}$(NO$_\mathrm{3}$)$_\mathrm{12}\cdot$24H$_\mathrm{2}$O (cerium magnesium nitrate), FAA~=~NH$_\mathrm{4}$Fe(SO$_\mathrm{4}$)$\cdot$12H$_\mathrm{2}$O (ferric ammonium alum), CPA~=~KCr(SO$_\mathrm{4}$)$\cdot$12H$_\mathrm{2}$O (chromium potassium alum) and MAS~=~Mn(NH4)$_\mathrm{2}$(SO$_\mathrm{4}$)$_\mathrm{2} \cdot$6H$_\mathrm{2}$O (manganese ammonium sulphate).}
\label{tab:ADR_materials}
\begin{indented}
\item[]\begin{tabular}{@{}llccc}
\br
Compound & S$_{\mathrm{GS}}$ & S$_{\mathrm{GS}}$/$\mathrm{vol.}$ & T$_{\mathrm{m}}$ & ADR T$_{\mathrm{min}}$ \\
 & & (mJ/K cm$^3$) & (mK) & (mK) \\
\mr
CMN \cite{Fisher1973} & Rln(2) & 16 & 2 &   \\
CPA \cite{Daniels1954} & Rln(4) & 42 & 10 &  \\
FAA \cite{Vilches1966} & Rln(6) &53 & 30 &  \\
KYbP$_{\mathrm{2}}$O$_{\mathrm{7}}$ \cite{Arjun2023Phosphate} & Rln(2) & 57 &  & 37 \\
KBaYb(BO$_{\mathrm{3}}$)$_{\mathrm{2}}$ \cite{Tokiwa2021,Jesche2023} & Rln(2) & 64	& 9 & 40 \\
NaYbP$_{\mathrm{2}}$O$_{\mathrm{7}}$ \cite{Arjun2023Phosphate} & Rln(2) & 64 &  & 45 \\
MAS \cite{Vilches1966} & Rln(6) & 70 & 170 &  \\
NaYbGeO$_{\mathrm{4}}$ \cite{Arjun2023} & Rln(2) & 101 & 210 & 135 \\
YbCu$_4$Ni \cite{Shimura2022}\cite{Sereni2018} & Rln(2) & 114 & 200 & 240 \\
YbPt$_{\mathrm{2}}$Sn \cite{Jang2015} & Rln(2) & 124 & 250 & 190 \\
Yb$_{\mathrm{3}}$Ga$_{\mathrm{5}}$O$_{\mathrm{12}}$ \cite{PaixaoBrasiliano2020} & Rln(2) & 124 & 180 & 100 \\
YbNi$_{\mathrm{1.6}}$Sn \cite{Gruner2024} & Rln(2) & 148 & 140 & 116 \\ 
YbBO$_{\mathrm{3}}$ \cite{Sala2023} & Rln(2) & 181 & 399 & 202 \\
KBaGd(BO$_{\mathrm{3}}$)$_{\mathrm{2}}$ \cite{Jesche2023} & Rln(8) & 192 & 263 & 122 \\
NaGdP$_{\mathrm{2}}$O$_{\mathrm{7}}$ \cite{NaGdP2O7} & Rln(8) & 210 & 570 & 215 \\
Gd$_{\mathrm{3}}$Ga$_{\mathrm{5}}$O$_{\mathrm{12}}$ \cite{Daudin1982,Kleinhans2023,Esat2021} & Rln(8) & 363 & 900 & 350 \\
GdPO$_{\mathrm{4}}$ \cite{Palacios2014} & Rln(8) & 401 & 725 & 385 \\
Gd$_{9.33}$[SiO$_4$]$_6$O$_2$ ~\cite{Yang2024} & Rln(8) & 509 & 500 & 300 \\
\br
\end{tabular}
\end{indented}
\end{table}

\begin{figure}[t]
\centering
\includegraphics[width=0.99\textwidth]{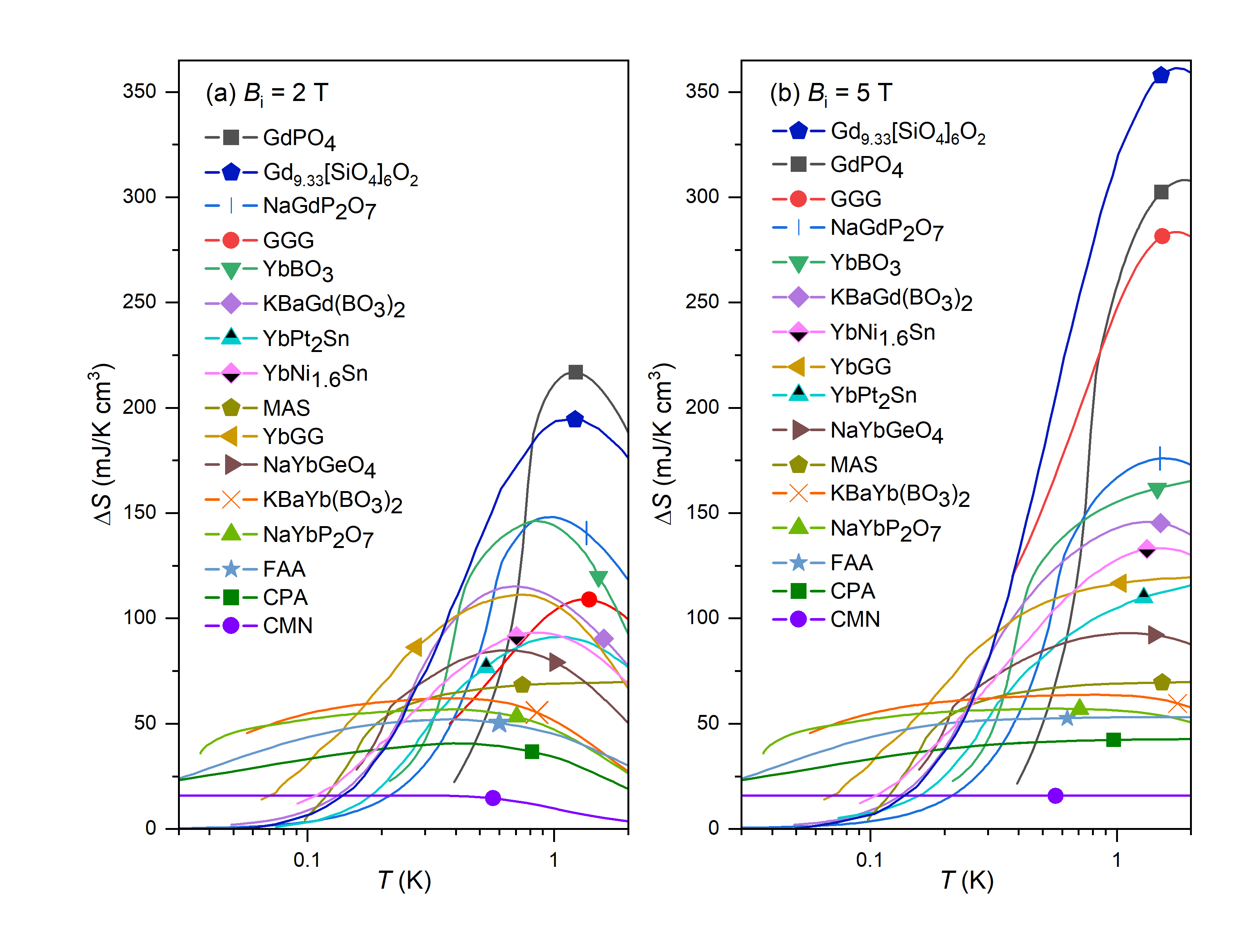}
\caption{Temperature dependence of the volumetric entropy density increment $\Delta S(T)=S(T,0)-S(T,B_i)$ at $B_i=2$~T (a) and $B_i=5$~T (b) for several mK ADR materials, calculated from available specific heat or entropy literature data cited in table \ref{tab:ADR_materials} (inter- and extrapolation were used occasionally).}
\label{fig:ADR_comparison}
\end{figure}

Next, the milli-Kelvin ADR performance of frustrated rare-earth oxides will be compared with that of other ADR materials, such as hydrated salts and Kondo metals.  Key parameters are listed in table \ref{tab:ADR_materials}. A large ground state entropy $S_{\mathrm{GS}}$ is advantageous, but it should be noted that for practical purposes the volumetric entropy density $S_{\mathrm{GS}}/{\mathrm{vol.}}$ is more relevant, listed in the third column.  Note that for the oxides presented, the Ag powder required for thermalization is not included in the values. The further two columns denote, as far as values are reported, magnetic ordering temperatures and experimentally obtained ADR $T_{\mathrm{min}}$ values. Of course the minimal temperature depends on the initial field and initial temperature as well as the non-perfect adiabatic conditions in various setups. The initial field and temperatures need to be chosen such that the entropy after isothermal magnetization is only a low fraction of the respective $S_{\mathrm{GS}}$. In the case of Yb- and Gd-oxides studied by us, $T_{\mathrm{min}}$ has been determined in the above described cooling experiments in the PPMS, starting at 2 K and 5~T. The non-adiabaticity of these experiments leads to a higher end temperature than theoretically possible. For instance, KBaYb(BO$_{\mathrm{3}}$)$_{\mathrm{2}}$ demagnetized under more adiabatic conditions in a dilution refrigerator yielded $T_{\mathrm{min}}$ below 20~mK. 

For KBaYb(BO$_{\mathrm{3}}$)$_{\mathrm{2}}$ and (Na,K)YbP$_{\mathrm{2}}$O$_{\mathrm{7}}$ almost all of the full ground-state entropy is used for cooling at an initial temperature of 2 K and an initial field of 5 T, which is also true for the listed paramagnetic salts CMN, CPA and FAA, while for KBaGd(BO$_{\mathrm{3}}$)$_{\mathrm{2}}$ a larger field would be necessary. In some applications also the required maximal field is a relevant parameter. Therefore, we calculated for figure~\ref{fig:ADR_comparison} the available entropy densities for 5~T (b) as well as a signifcantly lower initial field of 2~T (a).

In table \ref{tab:ADR_materials} the materials are sorted with increasing volumetric entropy densities. The upper materials, including three paramagnetic salts and three triangular lattice Yb-oxide magnets, allow for the cooling to temperatures significantly below 100 mK. In addition to the practical advantages, related to their chemical stability, the three listed frustrated Yb oxides also offer a higher entropy density compared to respective hydrated paramagnetic salts. Among the further listed Yb-oxides YbBO$_3$ offers the largest entropy density, which exceeds that of Yb$_{\mathrm{3}}$Ga$_{\mathrm{5}}$O$_{\mathrm{12}}$ (YbGG) and the different listed Yb-based intermetallic and heavy-fermion compounds. Due to its larger spin, Gd-oxides can offer higher entropy densities. There are numerous reports on Gd-based ADR materials in the literature, including molecular refrigerants~\cite{Tziotzi2023}, and we apologize for citing only a small fraction of them. To the best of our knowledge, among the Gd-based oxides, KBaGd(BO$_3$)$_2$ reaches the lowest ADR temperature of 122~mK. The highest entropy densities in our list are realized in Gd$_{\mathrm{3}}$Ga$_{\mathrm{5}}$O$_{\mathrm{12}}$ (GGG), GdPO$_4$ and most recently reported Gd$_{9.33}$[SiO$_4$]$_6$O$_2$ (GSO)~\cite{Yang2024}. In GSO the statistical distribution of 2/3 Gd vacencies leads to strongly broadened heat capacity maximum in zero field, centered around 500~mK. 

The general trend that higher entropy density, which enhances the cooling power for ADR, is counteracted by a higher minimal temperature, reported previously for paramagnetic salts~\cite{Wikus2014}, is overall also valid in table \ref{tab:ADR_materials}, though frustrated interactions help to keep $T_{\rm min}$ lower. In fact, frustrated magnets outperform respective paramagnetic salts in each temperature interval. This can be seen in figure~\ref{fig:ADR_comparison}. Following \cite{Gruner2024}, we plot therein the volumetric entropy density increment $\Delta S(T)=S(T,0)-S(T,B_i)$ at $B_i=2$~T (a) and $B_i=5$~T (b) for all materials listed in table \ref{tab:ADR_materials}. For either 2 or 5~T initial field, the entropy densities of the frustrated magnets clearly outperform that of paramagnetic salts in the entire temperature regime and in particular below 100 mK. From these plots, the material with the highest entropy change at a given temperature can be identified as the point of intersection between two curves, which makes one material a better refrigerant than the other for that temperature range. The desired lowest achievable temperatures must also be considered, as magnetic ordering limits the working range of materials with high entropy change. We note, that the minimal temperature and hold time of ADR cryostats for the temperature range below 100~mK can be optimized by pre-cooling thermal shielding and thermal busses for the wires to temperatures below 500 mK. This can be efficiently achieved by Gd-based frustrated magnets. Two- or multi-stage ADR has the important advantage, relevant in particular for space applications, that less magnetic field is required (sometimes below 2~T), since each stage cycles over only a fraction of the total operating range of a respective single-stage cooler~\cite{Shirron2014}.

%The compounds can be classified into two categories based on their volumetric entropy density and ordering temperatures. One group has a high volumetric entropy density and order at relatively high temperatures, making cooling below 100 mK impossible. These materials can be used as pre-cooling refrigerants to cool the primary cooling stage and all the connections leading to it. This reduces heat leakage and enables lower starting temperatures for further stages, allowing for cooling to even lower temperatures. The materials from the other group are relevant for this stage, as they allow for cooling well below 100 mK due to their very weak interactions, making them highly suitable for low-temperature applications. 

With respect to the highest entropy density increment GdPO$_{\mathrm{4}}$ and GSO reach the largest values for initial fields of 2 and 5~T, respectively, exceeding GGG. Below the AFM ordering in GdPO$_{\mathrm{4}}$ at 0.77 K \cite{Thiriet2005, Palacios2014} the entropy change decreases rapidly as shown in figure \ref{fig:MCE}. For an initial field of 5 T GGG outperforms GdPO$_{\mathrm{4}}$ (but not GSO) below 800 mK, because it only features short-range correlations, giving rise to a broad specific heat maximum near 0.9~K~\cite{Daudin1982}, followed by magnetic ordering at 350 mK at $\sim$ 1 T, preventing further cooling \cite{Schiffer1994}. However, for an initial field of 2 T, there are excellent alternatives, most prominently YbBO$_{\mathrm{3}}$, with the largest entropy density increments among Yb-based systems followed between 0.2 and 0.4 K by YbGG. Geometrical frustration lowers its ordering temperature to 54 mK. This has been discussed as a suitable candidate for space applications in the 200 mK - 2 K temperature range~\cite{PaixaoBrasiliano2020}. To achieve lower temperatures, KBaYb(BO$_{\mathrm{3}}$)$_{\mathrm{2}}$ and (Na,K)YbP$_{\mathrm{2}}$O$_{\mathrm{7}}$ are the most obvious options. As the curves for KYbP$_{\mathrm{2}}$O$_{\mathrm{7}}$ are very similar to those for NaYbP$_{\mathrm{2}}$O$_{\mathrm{7}}$, it has not been included in the graphs for visibility reasons. Due to the relatively high volumetric entropy density, all three materials outperform the established paramagnetic salts at temperatures below 200 mK.

\begin{figure}[t]
\centering
\includegraphics[width=0.99\textwidth]{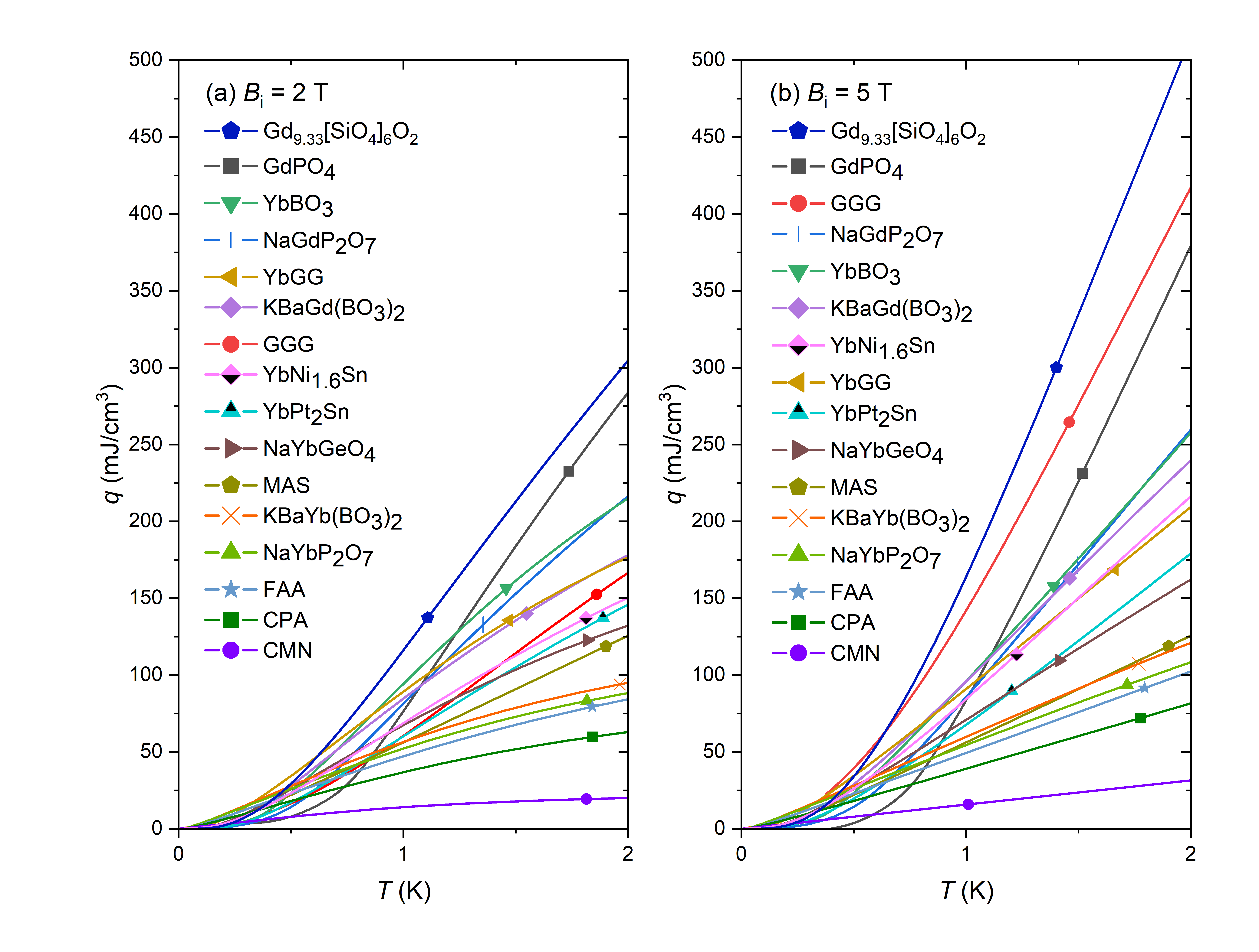}
\caption{Temperature dependence of the refrigerant capacity $q$ (see text), obtained by temperature integration of the respective data shown in figure~\ref{fig:ADR_comparison} starting at $\Delta S(T=0)=0$.}
\label{fig:refrigerant_comparison}
\end{figure}

As pointed out in~\cite{Gschneider2005}, the (volumetric) entropy density increment is not the only relevant criterion when comparing the suitability of ADR materials. We can use the temperature dependence of $\Delta S(T)$ to calculate the volumetric magnetic heat $q=\int_{T_1}^{T_2}\Delta S(T) dT$ that could be absorbed by the refrigerant material when warming from the cold temperature $T_1$ to the ''warm'' sink $T_2$ after demagnetization from $B_i$ to zero field. Maximizing this ''refrigerant capacity'' is desirable. Thus we used the data from figure~\ref{fig:ADR_comparison} to calculate $q(T)$, with $T$ being the upper integration limit. Here we started the numerical integration of the respective data at zero temperature by always adding one additional point $\Delta S(T=0)=0$. The respective curves are displayed in 
figure~\ref{fig:refrigerant_comparison}. For finding the best suitable material in a given temperature interval between $T_1$ and $T_2$, it is then required to compare the differences $q(T_2)-q(T_1)$ for the various materials.

%The high spin hydrated paramagnetic salts such as MAS and FAA, which have a larger magnetic entropy Rln(6), are also affected by their much higher magnetic ordering temperatures. On the other hand, low transition temperature materials such as CPA and CMN have a low magnetic moment density and hence a low entropy density. This shows that a high entropy density is generally incompatible with a low magnetic ordering temperature. NaYbGeO$_{\mathrm{4}}$ also suffers from magnetic ordering. The cooling power is only relevant in the intermediate temperature range, where it is larger than MAS. This again emphasises the importance of magnetic order suppressing effects such as frustration and site randomness in KBaYb(BO$_{\mathrm{3}}$)$_{\mathrm{2}}$.

\section{Summary and outlook}

ADR to milli-Kelvin temperatures provides a sustainable and scalable alternative to dilution refrigeration with $^3$He/$^4$He that could become important in the context of applications in quantum sensing and computing technology. Established hydrated paramagnetic salts have several limitations: most importantly, due to their chemical instability they must not be heated, making them inherently incompatible with UHV operation, and must be encapsulated to tolerate vacuum. In combination with their weak thermal conductivity this requires the elaborate  design of vacuum sealed ADR pills~\cite{Bartlett2014}. Ytterbium-based intermetallics have recently been proposed as suitable alternatives. In addition to their chemical stability and excellent thermal conductivity, facilitating a much simpler pill design, they also promise very large entropy density~\cite{Jang2015,Shimura2022,Gruner2024}. However, they generally have higher ordering temperatures than ionic compounds because of the Ruderman-Kittel-Kasuya-Yosida (RKKY) interaction~\cite{Wikus2014} and thus indeed cannot be used to cool to below 100 mK. While paramagnetic YbCo$_2$Sn$_{20}$ in principle allows lower ADR temperatures~\cite{Tokiwa2016}, its magnetic entropy is reduced even at 5~T by the Kondo effect, making this material less effective for mK-ADR ~\cite{Gruner2024}. The same holds true for Kondo lattices near a magnetic quantum critical point~\cite{Gegenwart2016}.

On the other hand, geometrically frustrated KBaYb(BO$_{\mathrm{3}}$)$_{\mathrm{2}}$ and (Na,K)YbP$_{\mathrm{2}}$O$_{\mathrm{7}}$ can provide very low ADR end temperatures (even down to 20 mK) in combination with a competitive volumetric entropy density. The importance of frustration is evident by comparison with non-frustrated NaYbGeO$_4$, which does not reach this temperature range, due to the formation of long-range ordering. In KBaYb(BO$_{\mathrm{3}}$)$_{\mathrm{2}}$, the ADR capabilities are likely further enhanced by the random distribution of K$^+$ and Ba$^{2+}$ ions, which can produce a non-uniform electric field leading to randomization of the magnetic couplings~\cite{Li2020}.

Our comparison of various materials with respect to the temperature dependent entropy density increment in figure~\ref{fig:ADR_comparison} allows to identify the best suitable system for a given working temperature up to 2~K. We note that the leading materials from this comparison feature magnetic frustration. However, little is known yet on the possible spin liquid phases in these ADR materials, owing to the fact that inelastic neutron scattering, preferentially on single crystals, would be required with very-high energy resolution for very-low energy transfer and at very-low temperatures, which is quite challenging. Interestingly, in this context, the comparison of thermodynamic data with theoretical modelling for various frustrated Gd-based ADR magnets revealed that the measured magnetocaloric cooling
rate down to 2~K results from two different contributions. One due to the number of soft modes, related to the magnetic frustration, superposed by the ordinary paramagnetic response, and the former one is expected to take over only at very low temperatures~\cite{Koskelo2023}.

In conclusion, this paper provides a comparison of recently studied mK-ADR materials with the particular focus on geometrically frustrated magnets. Several materials are identified, which outperform the classical hydrated paramagnetic salts and also offer important practical advantages. From the comparison between frustrated and non-frustrated magnetically ordered materials it is clear that lower ADR temperatures can be realized in the former ones. However, a more detailed microscopic understanding of frustration and (in some cases) structural randomness in these materials requires further detailed investigation.

The data associated with this paper can be found in Ref.~\cite{data}.

%In addition, more research is needed on high magnetocaloric materials, as their potential for ADR has not yet been fully exploited. One possible approach is to investigate high magnetic density systems where dipolar, exchange and anisotropy interactions compete to prevent ordering at low temperatures. Another important factor is to provide further support for the large-scale application of frustrated magnets in real devices, potentially replacing dilution refrigeration and enabling ultra-low temperatures to be achieved in a more sustainable and affordable way. Commercial continuous refrigerators based on ADR are available on the market, but they are still limited in terms of materials and technology. Compared to dilution refrigerators, greater cooling capacities and faster cooling rates may be achievable. Further research is needed to establish ADR as a dominant technique for achieving ultra-low temperatures. Frustrated magnetism is one of the most active areas of research in modern condensed matter physics, and it will be fascinating to see what progress is made in this field in the near future.

\ack
We would like to thank U. Arjun, S. Bachus, A. Bellon, F. Hirschberger, K. Kavita, K.M. Ranjith, D.D. Sarma, Y. Tokiwa and A.A. Tsirlin for collaborative work. Stimulating discussions within the CNRS international research network on strongly correlated electron systems as advanced magnetocaloric materials, led by A. Honecker, are gratefully acknowledged. This work was supported by the German Research Foundation (DFG) through Project 514162746 (GE 1640/11-1).
\section*{References}
\bibliography{Literature_IOP_ADR_v2}

\begin{thebibliography}{10}
\bibitem{Xac}
The measurement was done using a home-built setup with a customised pair of
  anti-Helmholtz pick-up coils and a modulation coil located on the outside of
  the inner vacuum chamber. The measured induced voltage was amplified and
  measured using a lock-in amplifier, which is also used for the output of the
  modulation coil. To obtain absolute values of the susceptibility, the low
  temperature AC susceptibility data are scaled to the DC susceptibility data
  measured with a Quantum Design MPMS.
\end{thebibliography}

\end{document}